# Hole transport across MgO-based magnetic tunnel junctions with high resistance-area product due to oxygen vacancies

F. Schleicher[1,2], B. Taudul[1], U. Halisdemir[1], K. Katcko[1], E. Monteblanco[2], D. Lacour[2], S. Boukari[1], F. Montaigne[2], E. Urbain[1], L. M. Kandpal[1], J. Arabski[1], W. Weber[1], E. Beaurepaire[1†], M. Hehn[2], M. Alouani[1], M. Bowen[1,*]

[1]IPCMS UMR 7504 CNRS, Université de Strasbourg, 23 Rue du Loess, BP 43, 67034 Strasbourg Cedex 2, France

[2]Institut Jean Lamour UMR 7198 CNRS, Université de Lorraine, BP 70239, 54506 Vandoeuvre les Nancy, France

†: deceased April 24th, 2018.

**Abstract:**

The quantum mechanical tunnelling process conserves the quantum properties of the particle considered. As applied to solid-state tunnelling (SST), this physical law was verified, within the field of spintronics, regarding the electron spin in early experiments across Ge tunnel barriers, and in the 90s across $Al_2O_3$ barriers. The conservation of the quantum parameter of orbital occupancy, as grouped into electronic symmetries, was observed in the '00s across MgO barriers, followed by $SrTiO_3$ (STO). Barrier defects, such as oxygen vacancies, partly conserve this electronic symmetry. In the solid-state, an additional subtlety is the sign of the charge carrier: are holes or electrons involved in transport? We demonstrate that SST across MgO magnetic tunnel junctions (MTJs) with a large resistance-area (RA) product involves holes by examining how shifting the MTJ's Fermi level alters the ensuing barrier heights defined by the barrier's oxygen vacancies. In the process, we consolidate the description of tunnel barrier heights induced by specific oxygen-vacancy induced localized states. Our work opens prospects to understand the concurrent observation of high TMR and spin transfer torque across MgO-based nanopillars.

To discriminate between electron and hole transport in a metallic system, one examines the curvature of the energy dispersion curve at the Fermi level. Conversely, electron(hole) tunnelling in the solid-state occurs with respect to energy barrier heights defined as the energetical separation between available (occupied) electronic states and the Fermi level [1]. The most obvious source of such states are the extended states of the conduction(valence) bands of the tunnel barrier's dielectric, which can define electron(hole) barrier heights $\phi_e$ and $\phi_h$ for carriers whose spin and electronic symmetry match those of the band considered.

The complexity of spin, symmetry, and charge carrier-resolved SST was nicely encapsulated through measurements [2] of fully spin- and symmetry-polarized hole transport from $La_{1-x}Sr_xMnO_3$ (LSMO) electrodes through a $SrTiO_3$ transition metal oxide (TMO) tunnel barrier with a ~ 3 eV band gap and $\phi_e < \phi_h$. Indeed, a barrier height that exceeded half of the STO band gap was observed experimentally and attributed to $\phi_h$, consistently with a matching of electronic symmetry between the the LSMO $\Delta_1^\uparrow$ charge carriers and the STO valence band with partial $\Delta_1$ character. The absence of any conductance increase upon reaching $\phi_e$ reflects a symmetry mismatch between the STO conduction band's $\Delta_{2',5}$ character and the LSMO $\Delta_1^\uparrow$ charge carriers. Due to electronic symmetry, holes are tunnelling across this electron tunnel junction.

The vast majority of SST-based spintronic studies utilize, however, transition metal electrodes that do not exhibit completely spin-/symmetry-polarized transport. Furthermore, they are associated with ionic oxide barriers, such as $Al_2O_3$ or MgO, whose valence and conduction bands share the $\Delta_1$ electronic symmetry [3]. Reports utilizing either phenomenological free-electron models (e.g. Brinkman or Simmons) [4,5], or more physical models such as the $\hat{i}$ method [2,6–9], have identified barrier heights across MgO that are systematically much lower than the expected 4 eV arising from extended states [10,11]. The resulting high current densities in turn enable tunnelling spin transfer torque physics [12], with application ranging from information encoding [13] to neuromorphic computing [14]. A set of recent reports [8,9] highlight how the localized states induced by oxygen vacancies in MgO [15] can, by virtue of their electronic properties, lower $\phi$ while promoting a level of spintronic performance that can only be explained in terms of conservation of the electron's spin and orbital occupancy. Further improvements to spintronic performance can be achieved by altering the MgO barrier growth and annealing procedure [9,15,16], so as to increase the ratio of double (denoted M centers) to single (denoted F centers) oxygen vacancies [9].

Ab-initio theory quantitatively attributes the experimentally observed barriers height to the occupied ground state of these single and double vacancies [9]. In this Letter, we confirm that holes are tunnelling with respect to these occupied states by examining changes to the potential landscape induced by oxygen vacancies upon shifting the MTJ's Fermi level. We experimentally observe an increase in the barrier heights defined by the (occupied) ground states of F and M centers upon replacing the $Fe_{0.4}Co_{0.4}B_{0.2}$ (FeCoB) top electrode with $Fe_{0.8}B_{0.2}$ (FeB) that is in reasonable quantitative agreement with ab-initio calculations. We propose a revised diagram of the potential tunnelling landscape induced by oxygen vacancies.

To shift the Fermi level of our FeCoB/MgO/FeCoB MTJs, we replaced the FeCoB top electrode with FeB. By maintaining the FeCoB composition of the lower electrode, we ensure that no change in the MgO barrier's structure occurs during growth [17]. We've also confirmed that, for the MgO barriers sputtered using a MgO target of the present work, annealing has no impact on the nature/density of localized states as determined from magnetotransport [16]. It is therefore reasonable to conclude that the structural properties of the MgO barriers in MTJ stacks with FeCoB and FeB top electrodes are comparable.

Ta(5)/Co(10)/IrMn(7.5)/FeCoB(4)/MgO(2.5)/FM(4.5)/Ta(5)/Pt(5) stacks (FM is either FeCoB or FeB; thicknesses are in nm) were grown on Corning 1737 substrates by standard sputtering techniques. A subsequent 300 ºC annealing step in an external magnetic field sets the exchange field of the IrMn pinning layer within the lower electrode and promote the diffusion of B to crystallize both FeCoB and FeB magnetic electrodes [18]. Details of the growth may be found in previous reports [8,19]. MTJs with a diameter of 20 μm were then defined using standard photolithography techniques [8,20]. Typical RA products obtained are $5*10^7$ $\Omega.\mu m^2$ ($1*10^8$ $\Omega.\mu m^2$) for the CoFeB (FeB) sample. Devices were contacted in 4-point mode with positive(negative) contacts on the top(bottom) electrode, such that holes are flowing from top to bottom at positive bias voltage. Typical TMR values are 105 % at room temperature (for both types of samples) and 205 % (160 %) for FeCoB (FeB) at 10 K. The difference is not surprising, due to weaker symmetry filtering at the Fe/MgO interface [21,22].The following results are representative of previous studies on FeCoB/MgO/FeCoB MTJs [8,16], which we compare with results obtained on six FeCoB/MgO/FeB MTJs.

To witness the presence and spintronic impact of localized states, we examine the impact $TMR_{rel}(V,T)$ on the bias dependence of tunnel magnetoresistance (TMR) of increasing temperature in 9 K temperature intervals, i.e. $TMR_{rel}(V,T)=[[TMR(V,T)/TMR(V,T-9K)]-1]*100$ (%), where $TMR(V,T_0)=I_P(V,T_0)/I_{AP}(V,T_0)-1$. This approach constitutes a spintronic implementation of the $Î(V,T_1)=I(V,T_1)/I(V,T_0)-1$ method [6–8], which attributes the bias position of maxima in $Î$ to the energetical distance from $E_F$ of a DOS feature that has been thermally activated upon increasing T from $T_0$ to $T_1$. In contrast to phenomenological models (e.g. Simmons or Brinkman) [4,5], this physical method of determining features in the MTJ potential landscape makes no assumption regarding the effective mass or energy dispersion of the charge carrier [7].

We present in Fig. 1 $TMR_{rel}(V,T)$ for a MTJ with (a) FeCoB top electrode, and (b) a FeB top electrode. The 40 K < T < 300 K temperature range accessible here does not extend low enough to reveal the intrinsic features of the electrodes, but can probe defect-mediated potential landscape features [8]. Referring to panel (a), when two FeCoB electrodes are used, the TMR thermal decrease at 50 K is largest at 1.2 eV, and decreases within 50 K < T < 100 K. Although this was interpreted as arising from the single oxygen vacancy $F/F^+$ (i.e. neutral/singly-charged) ground state located 1.2 eV below $E_F$, later ab-initio calculations pegged this state to be the bonding ground state $M_1$ of the double oxygen vacancy [9]. Furthermore, the maximum TMR thermal decrease at 0.7 V for 100 K < T < 250 K was attributed [8] to the excited state of the charged single oxygen vacancy $F^+$, denoted $F^{+*}$ and necessarily above $E_F$. We will experimentally and theoretically demonstrate in what follows that this is in fact the single oxygen vacancy's ground state F, in agreement with theory [9]. Finally, the further decrease in the $TMR_{rel}$ maximum to ~ 0.4 V for T > 250 K was attributed [8,16] to the double oxygen vacancy's antibonding ground state $M_2$ [9].

Comparing now panels (b) and (a) of Fig. 1, when the top electrode is switched from FeCoB to FeB, we find that the entire set of features shifts by 0.45 eV to higher energies. This is especially visible for the states initially located at 0.4 eV and 0.7 eV, while the shift of the state initially at 1.2 eV is inferred from the rise in $TMR_{rel}$ amplitude at the edge of our 1.6 V-wide spectroscopic dataset [23]. Within experimental error (of 0.04 eV for CoFeB and 0.1 eV for FeB datasets), we do not observe a concurrent asymmetry in energy position of the localized states as would be expected when probing the FeCoB/MgO and MgO/FeB interfaces with holes at positive and negative applied bias. This suggests that the underlying assumptions of (i) a rigid energy band model of the tunnel junction; of (ii) a uniform spatial distribution of oxygen vacancies across the barrier; and of (iii) a charge carrier tunnelling from the injecting electrode to the collecting electrode, while at the state-of-the-art, may need to be refined in future research.

To confirm that this trend reflects changes in barrier height induced by oxygen vacancies due to a modification in the Fermi level position within the MgO band gap, we present in Fig. 2 ab-initio calculations of the F and M centers' occupied states within MgO (001) as the adjoining electrode is switched from *bcc* Fe (001) to FeCo (001) with either a pure Co interface or a FeCo interface. The F and M centers are located on the center monolayer of a seven-monolayer thick MgO layer sandwiched between either Fe, FeCo or Co electrodes [24]. We used the projector augmented plane wave method as implemented in VASP code within the generalized gradient approximation (GGA) with Pedrew, Burke and Erzerhof parametrization. Although the MgO band gap is underestimated by GGA, the energy position of the localized states relative to the electrode Fermi level is quantitatively correct as demonstrated [9] by comparing with calculations performed using the computationally more intensive HSE03 (Heyd-Scuseira-Ernzerhof) hybrid functional [25], which yields the correct energetical band gap.

Referring to Fig. 2, when Fe electrodes are used instead of FeCo, we observe that the F center energy position E shifts from -0.9 eV to -1.15 eV (panel (a)), while the $M_1$/$M_2$ states shift from -1.42 eV/-0.48 eV to -1.63 eV/-0.71 eV (panel (b)). Thus, we find that the MTJ's Fermi level moves away from the F and M center ground states by 0.25 eV, in reasonable quantitative agreement with the experimentally observed shift in the tunnel barrier height inferred from the experimental data of Fig. 1. As a note, for calculations involving MgO sandwiched between Co electrodes we find that the Fermi level moves closer by 0.15 eV. Results are summarized schematically on panel c. Note how we also obtain a reasonable agreement between experiment and theory regarding the energy position of the states. Indeed, the only minor discrepancy regards the 1.2 eV and 1.48 eV energy positions of the $M_1$ state below $E_F$ that are respectively found experimentally and theoretically.

While prior experimental results were interpreted in terms of hole tunnelling with respect to occupied states [8,9,12,16], the present experimental/theoretical datasets constitute the first evidence that hole tunnelling is taking place in MgO-based MTJs that combine low (< 3.9 eV) tunnel barrier heights and high (~$10^8$ $\Omega.\mu m^2$) RA products. As a note, this identification of localized states is compatible with two previous reports. First, within conducting tip atomic force microscopy experiments on Fe/MgO bilayers [15], this attributes hotspots on a MgO barrier grown without(with) oxygen in the sputter plasma to $M_2$(F) centers around 0.5 eV (1.1 eV) away from $E_F$, while the corresponding background corresponds to F centers (extended band states) at 1.1 eV and ~ 4 eV from $E_F$. Second, within magnetotransport across FeCoB/MgO-class MTJs in which the MgO layer is grown from post-oxidized Mg, increasing the post-annealing temperature converts oxygen vacancies from F to M centers with corresponding barriers heights decreasing from ~ 0.8 eV to ~ 0.4 eV [9].

In light of the present results, we reassess our previously reported [8] correlation between the temperature dependencies of photoluminescence (PL) and magnetotransport in terms of these localized states. To explain the correlation, we proposed that reduced localization upon temperature increase would both reduce the PL by competing against the carrier recombination time and, by promoting hopping transport, enhance the signature of a tunnel barrier in î maps. While this physical picture of the correlation likely remains valid, several aspects impede a direct link between these optical and electrical mechanisms to populate/depopulate the localized states of oxygen vacancies. First, our present understanding of excited states, both experimentally and theoretically, remains inadequate. Second, while the origin of the 2.2 - 2.3 eV, 2.6 eV and 3.5 eV PL emission lines appears to be resolved [26], there are no reports of the additional emission line that was experimentally observed [8] at 1.7 eV. Resolving these aspects requires future PL experiments on MgO in which the ratio of F to M centers can be nominally varied [9].

To conclude, we combined magnetotransport experiments and ab-initio theory to show that holes predominantly tunnel across MgO barriers in the presence of isolated single and/or double oxygen vacancies. To do so, we shifted the Fermi level by replacing FeCoB with FeB as our MTJs' top electrode. We find that all localized states shift to a higher bias value. Ab-initio calculations show that moving from Co to FeCo to Fe interfaces yields a similar increase in energy separation between the ground states of the single (F center) and double (M center) oxygen vacancies away from the Fermi level. The amplitude of the experimentally and theoretically determined shifts is in reasonable quantitative agreement, as is the energy position of these localized states. We are thus able to identify the oxygen vacancies in previous reports with a high degree of certainty across systems with varying electrode materials.

Looking ahead, our work opens prospects to address the concurrent observation, in MgO-based MTJ nanopillars, of large TMR and very low barrier heights leading to sufficient current densities to observe the spin transfer torque (STT) effect [14]. Indeed, these nanopillars typically exhibit a RA~1-10 $\Omega.\mu m^2$ across a ≈1nm-thick MgO barrier [27]. Yet, at this ~5ML thickness, a very low barrier height is required to enable (STT) [12], while a metal-induced gap states picture alone cannot attenuate the band gap on the barrier's central MgO monolayer since it is shielded by two intervening MgO monolayers [28]. We will discuss these issues through experiments on academic and industrial nanopillars in a future paper.

**Acknowledgements:**

We acknowledge funding from the Agence Nationale de la Recherche (ANR-09-JCJC-0137, ANR-14-CE26-0009-01), CEFIPRA (5604-3) and the Labex NIE "Symmix" (ANR-11-LABX-0058 NIE). Devices were synthesized at the STNano technological platform.

**Figures and Captions**

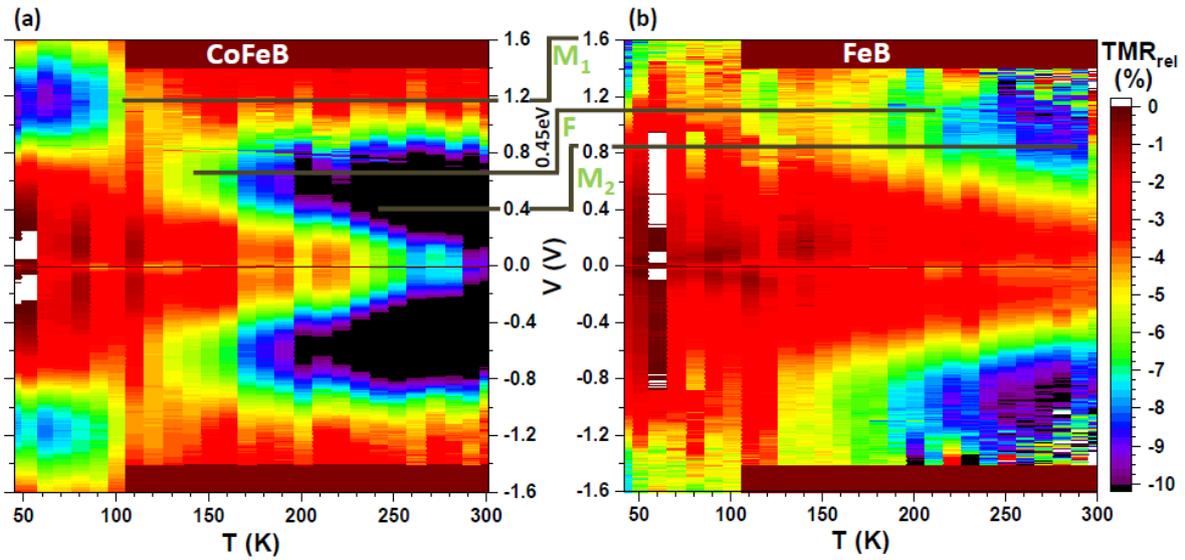

Figure 1: **Bias shift of localized states within thermal impact on magnetotransport upon varying the MTJ counterelectrode**. Impact TMR$_{rel}$(V,T) on the bias and temperature dependence of TMR of increasing temperature, in 9 K temperature intervals, on MgO MTJs with a (a) FeCoB and (b) FeB top electrode. A 0.45 eV shift is materialized by straight lines straddling both panels that pinpoint the common, but bias-shifted, feature arising from the $M_1$, F and $M_2$ centers. Higher level of noise in the FeB dataset visible in the high (V,T) area results from dividing two very low TMR values, lower than in the CoFeB dataset.

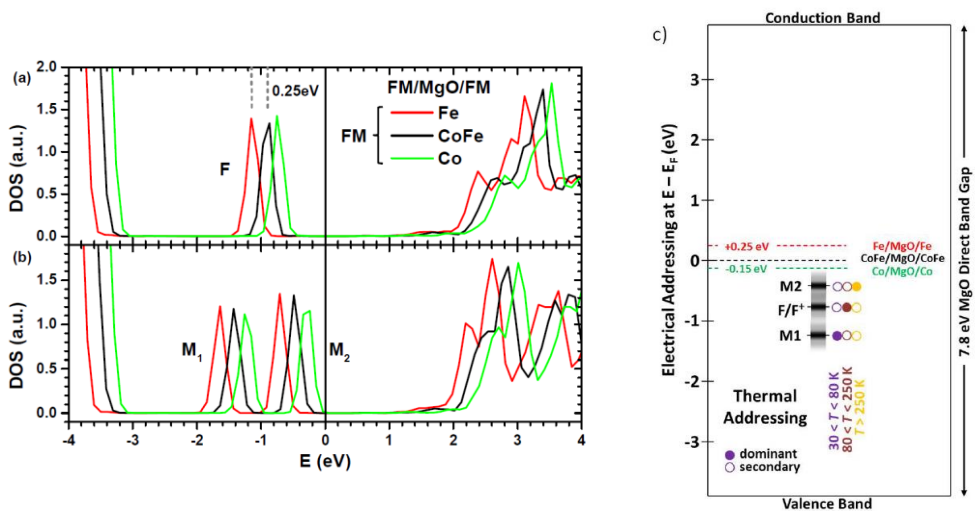

Figure 2: **Altering the energy separation of oxygen vacancy states in MgO from a MTJ's Fermi level by varying the MTJ electrode.** Ab-initio densities of states of MgO (001) comprising (a) F and (b) M centers, calculated for a MTJ stack as the interface is switched from *bcc* Fe (001) (black) to FeCo (001) (red) to Co (001) (green). (c) Schematic diagram of the energy positions of localized states due to oxygen vacancies in the MgO barrier, relative to the electrode-defined Fermi level. Circles denote the temperature range at which a localized state experimentally dominates the tunnelling potential landscape.